\documentclass[11pt]{article}
\usepackage{amssymb,amsmath,amsthm,amsfonts,amscd,latexsym}
\usepackage{mathrsfs}
\usepackage{graphicx}
\usepackage{epstopdf}

\renewcommand{\paragraph}{\roman{paragraph}}
\input cyracc.def

\newcommand{\Z}{\mathbb{Z}}

\begin{document}
\title{\bf On constacyclic codes over $\mathbb{Z}_4[u]/\langle u^2-1\rangle$ and their Gray images}
\author{{Minjia Shi\thanks{Corresponding author: Minjia Shi, Key Laboratory of Intelligent Computing Signal Processing, Ministry of Education, Anhui University, No.3 Feixi Road, Hefei, Anhui, 230039, China, School of Mathematical Sciences, Anhui University, Hefei, Anhui, 230601,
China and National Mobile Communications Research Laboratory, Southeast University, China. E-mail: smjwcl.good@163.com.}, Liqin Qian\thanks{Liqin Qian, School of Mathematical Sciences, Anhui University, China. E-mail: qianliqin$\_1108$@163.com.}, Lin Sok\thanks{Lin Sok, School of Mathematical Sciences, Anhui University, Hefei, Anhui, 230601 and Department of Mathematics, Royal University of Phnom Penh, Cambodia. E-mail: sok.lin@rupp.edu.kh.}, Nuh Aydin\thanks{Nuh Aydin, Department of Mathematics and Statistics, Kenyon College, Gambier, OH, 43022, United States, E-mail: aydinn@kenyon.edu}, Patrick Sol\'{e}\thanks{Patrick, Sol\'{e}, CNRS/ LTCI, University Paris-Saclay, 75013 Paris, France, E-mail: patrick.sole@telecom-paristech.fr}}
}
\date{}
\maketitle

{\bf Abstract:} {\normalsize  We first define a new Gray map from $R=\mathbb{Z}_4+u\mathbb{Z}_4$ to $\mathbb{Z}^{2}_{4}$, where $u^2=1$ and study $(1+2u)$-constacyclic codes over $R$. Also of interest are some properties of $(1+2u)$-constacyclic codes over $R$. Considering their $\mathbb{Z}_4$ images, we prove that the Gray images of $(1+2u)$-constacyclic codes of length $n$ over $R$ are cyclic codes of length $2n$ over $\mathbb{Z}_4$.  In many cases the latter codes have better parameters than those in the online database of Aydin and Asamov. We also give a corrected version of a table of new cyclic $R$-codes  published by \"Ozen et al. in Finite Fields and
Their Applications, {\bf 38}, (2016) 27-39. }


\section{Introduction}
\hspace*{0.5cm}
Codes over finite rings have been studied since the
early 1970s. A great deal of attention has been given to codes over
finite rings since the middle of 1990s because of their new role in algebraic
coding theory and their useful applications. A landmark paper \cite{94}
has shown that certain nonlinear binary codes with excellent
error-correcting capabilities can be identified as images of linear
codes over $\mathbb{Z}_4$ under the Gray map. This motivated the study of codes over finite rings, especially codes over $\mathbb{Z}_4$, which remain a special topic of interest in the field because of their relation to lattices, designs, and low correlation sequences \cite{W}.

Due to the importance of codes over $\mathbb{Z}_4$ and intensive work over $\mathbb{Z}_4$-codes, a database of $\mathbb{Z}_4$-codes was created in \cite{A} and is available in \cite{D}. An important problem in the field is to obtain codes over $\mathbb{Z}_4$ with better parameters than the ones given in the database \cite{D}. A lot of work for this has been done in recent years (e.g.\cite{A,22,MJ,wg,yn,yk}). For example, Wu et al discussed $1$-generator generalized quasi-cyclic codes over $\mathbb{Z}_4$ in \cite{wg}. They constructed some new $\Z_4$-linear codes and obtained some good binary nonlinear codes using the usual Gray map.

More recently, extension rings of $\mathbb{Z}_4$ have been considered in coding theory. Among  those, rings of order $16$ are of special importance \cite{MS}. Yildiz, Aydin and Karadeniz discussed linear codes, cyclic codes over $\mathbb{Z}_4+u\mathbb{Z}_4$ $(u^2=0)$ and $\Z_4$-images in \cite{yn, yk}. The construction of one-Gray weight and two-Gray weight codes over $\mathbb{Z}_4+u\mathbb{Z}_4$ with $u^2=u$ was studied in \cite{MJ}. It is worth noting that \"{O}zen et al. have proved that the $\mathbb{Z}_4$-image of a $(2+u)$-constacyclic code over $\mathbb{Z}_4+u\mathbb{Z}_4$ ($u^2=1$) of odd length is a cyclic code over $\mathbb{Z}_4$ in \cite{22}. They also presented many examples of cyclic codes over $\mathbb{Z}_4+u\mathbb{Z}_4$ whose $\mathbb{Z}_4$-images have better parameters than previously best-known $\mathbb{Z}_4$-linear codes. Checking the parameters of the new codes presented in \cite{22}, we noted that the entries in Table 1 of \cite{22} are erroneous. One of the authors of \cite{22} (who is also a co-author of this paper) confirmed that somehow incorrect data  was entered into Table 1. We present the correct data in this paper.

We ask the following question to extend the work in \cite{22}. Does there exist a special class of constacyclic codes over $\mathbb{Z}_4+u\mathbb{Z}_4$ ($u^2=1$) whose $\mathbb{Z}_4$-images (with possibly a different Gray map) produce $\mathbb{Z}_4$-cyclic codes with improved parameters ? We have been able to show that the answer is affirmative. Using $(1+2u)$ as the shift constant, and a new Gray map we introduce, we have obtained many examples of constacyclic codes over $R$ whose $\mathbb{Z}_4$-images have better parameters than previously best-known $\mathbb{Z}_4$-linear codes given in \cite{D}.

The material of the paper is organized as follows. Section 2 introduces some preliminary results on linear codes over the ring $R$ that we need. In Section 3, we investigate the structures and properties of $(1+2u)$-constacyclic codes over $R$. In Section 4, we present  some cyclic codes over $\mathbb{Z}_4$ that are  obtained from the $(1+2u)$-constacyclic codes over $R$, and  have either the same parameters as the ones in \cite{D} or better parameters.  Section 5 concludes the paper.

\section{Preliminary results}

\hspace*{0.5cm}Throughout this paper, we let $R$ denote the commutative ring $\mathbb{Z}_4+u\mathbb{Z}_4=\{0, 1, 2, 3, u, 2u,\\ 3u, 1+u, 2+u, 3+u, 1+2u, 2+2u, 3+2u, 1+3u, 2+3u, 3+3u\}$, where $u^2=1$. Clearly, $R\cong\mathbb{Z}_4[u]/\langle u^2-1\rangle.$ Its units are given by $\{1, 3, u, 3u, 2+u, 1+2u, 3+2u, 2+3u\}.$ There are $7$ ideals in this ring of characteristics $4$ given by $\{\langle0\rangle, \langle2u\rangle, \langle1+u\rangle, \langle3+u\rangle, \langle2+2u\rangle, \langle2u, 1+u\rangle, R\}.$ It is a local ring with maximal ideal $\langle2u, 1+u\rangle$.

Let $\eta$ be a unit in $R$. A linear code $C$ of length $n$ over $R$ is called $\eta$-constacyclic if it is invariant under the constacyclic shift operator $\varrho_{\eta}(r_0, r_1,\cdots, r_{n-1}) = (\eta r_{n-1}, r_0, r_1, \cdots, r_{n-2}),$ where $(r_0, r_1,\cdots, r_{n-1})\in C$. The constant $\eta$ is called the shift constant for $C$. In this paper, we study constacyclic codes with shift constant $1+2u$ over $R$. Note that cyclic codes are a special case of constacyclic codes with $\eta=1$. If $\sigma$ is the cyclic shift operator, then $\sigma(r_0, r_1,\cdots, r_{n-1}) = (r_{n-1}, r_0, r_1, \cdots, r_{n-2}).$ In other words, $C$ is said to be cyclic if $\sigma(C)=C$ and constacyclic if $\varrho_{\eta}(C)=C$ for some unit $\eta\in R.$

Let $x=(x_0,x_1, x_2, \cdots, x_{n-1})$ and $y=(y_0,y_1, y_2, \cdots, y_{n-1})$ be two elements of $R^n$. The Euclidean inner product of $x$ and $y$ is defined as $x\cdot y=x_0y_0+x_1y_1+\cdots+x_{n-1}y_{n-1},$ where the operations are performed in $R$. For a code $C$ over $R$, its dual code $C^\bot$ is defined as $C^\bot=\{x\in R^n\mid x\cdot y=0$
for all $y\in C$\}.

Throughout this paper, we choose the unit $\eta=1+2u\in R$ as the shift constant of the constacyclic codes. It is well known that a $(1+2u)$-constacyclic code of length $n$ over $R$ can be identified as an ideal of the quotient ring $\frac{R[x]}{\langle x^n-(1+2u)\rangle}$ via the $R$-module isomorphism $\varphi:$
$$R^n\longrightarrow \frac{R[x]}{\langle x^n-(1+2u)\rangle},$$
\hspace*{3.1cm}$(a_0, a_1, \cdots, a_{n-1})\longmapsto a_0+a_1x+\cdots+a_{n-1}x^{n-1}({\rm mod} \ ( x^n-(1+2u))).$

In the sequel, we define a Gray map $\Phi: R\longrightarrow \mathbb{Z}_4^2$ by $\Phi(a+bu)=(b, 2a+b).$ One can verify that $\Phi$ is a linear map, but it is not a bijective map.

The polynomial correspondence of the Gray map can be defined as $$\Phi: R[x]/\langle x^n-(1+2u)\rangle\longrightarrow \mathbb{Z}_4[x]/\langle x^{2n}-1\rangle$$ given by $$\Phi(a(x)+b(x)u)=b(x)+x^n (2a(x)+b(x)).$$

Notice that $(1+2u)^n=1+2u$ if $n$ is odd and $(1+2u)^n=1$ if $n$ is even.
\section{$(1+2u)$-constacyclic codes over $\mathbb{Z}_4[u]/\langle u^2-1\rangle$}

\hspace*{0.5cm}Similarly to the proof of Proposition 4.1 and Theorem 4.2 in \cite{22}, we have the following proposition.\\
{\bf Proposition 3.1} Let $\varrho$ denote the $(1+2u)$-constacyclic shift of $R^n$ and $\sigma$ denote the cyclic shift of $\mathbb{Z}_4^n$. If $\Phi$ is the Gray map from $R^n$ into $\mathbb{Z}_4^{2n},$ then $\Phi\varrho=\sigma\Phi.$\\

As a consequence of Proposition 3.1, we have the following corollary.\\
{\bf Corollary 3.2} The Gray image of a $(1+2u)$-constacyclic code over $R$ of length $n$ is a cyclic code over $\mathbb{Z}_4$ of length $2n$.\\

\hspace{-0.6cm}{\bf Proposition 3.3}
Let $C$ be a code of length $n$ over $R$. Then $C$ is a $(1+2u)$-constacyclic code if and only if $C^\bot$ is a $(1+2u)$-constacyclic code.
\begin{proof} $\Longrightarrow:$ Let $C$ be a $\lambda$-constacyclic code of length $n$ over $R$ where $\lambda=1+2u$, and $x\in C^\bot$, $y\in C$. Because $C$ is $\lambda$-constacyclic, $\varrho_{\lambda}^{n-1}(y)\in C$, where the exponent $n-1$ denotes an $(n-1)$-fold composition. Thus, $0=x\cdot \varrho_{\lambda}^{n-1}(y)=\lambda\varrho_{{\lambda}^{-1}}(x)\cdot y=\varrho_{{\lambda}^{-1}}(x)\cdot y$, which means that $\varrho_{{\lambda}^{-1}}(x)\in C^\bot$. Therefore, $C^\bot$ is closed under the $\varrho_{{\lambda}^{-1}}$-shift. Since $\lambda^{-1}=(1+2u)^{-1}=1+2u$, $C^\bot$ is a $(1+2u)$-constacyclic code.\\
\hspace*{0.5cm}$\Longleftarrow:$ Suppose $C^\bot$ is a $(1+2u)$-constacyclic code. By the $``\Longrightarrow"$ direction,  $(C^\bot)^\bot$ is also a $(1+2u)$-constacyclic code.
\end{proof}
In some occasions, we find it is more convenient to use a permuted version of $\Phi_\pi$ defined as $\Phi_\pi(r)=(b_0, 2a_0+b_0, b_1, 2a_1+b_1,\cdots, b_{n-1}, 2a_{n-1}+b_{n-1}).$ The codes obtained using $\Phi$ and $\Phi_\pi$ are permutation equivalent.\\
{\bf Proposition 3.4}
For any $r\in R^n$, we have $\Phi_\pi\sigma(r)=\sigma^2\Phi_\pi(r).$
\begin{proof} The proof is similar to that of Proposition 4.3 in \cite{22}. We omit it here.
\end{proof}
From the definition of $\Phi_\pi$, we  get the following corollary.\\
{\bf Corollary 3.5}
Let $C$ be a cyclic code of length $n$ over $R$. Then its $\mathbb{Z}_4$-image $\Phi_\pi(C)$ is equivalent to a $2$-quasicyclic code of length $2n$ over $\mathbb{Z}_4$.\\

In the following, we study $(1+2u)$-constacyclic codes over $R$ when $n$ is odd by introducing the following isomorphism from $R_n$ to $T_n$.\\
{\bf Proposition 3.6}
Let $$\phi: R_n=R[x]/(x^n-1)\longrightarrow T_n=R[x]/(x^n-(1+2u))$$ be defined by $\phi(c(x))=c((1+2u)x).$
If $n$ is odd, then $\phi$ is a ring isomorphism.
\begin{proof} The proof is similar to that of Proposition 4.5 in \cite{22}. We omit it here.
\end{proof}
From Proposition 3.6, we obtain the following corollaries.\\
{\bf Corollary 3.7}
Let $n$ be an odd number. Then $I$ is an ideal of $R_n$ if and only if $\phi(I)$ is an ideal of $T_n.$

\hspace*{-0.6cm}{\bf Corollary 3.8} Let $n$ be an odd number. Then a $(1+2u)$-constacyclic code of length $n$ over $R$ is equivalent to a cyclic code of length $n$ over $R$ by the ring isomorphism $\phi$.\\

Let $\overline{\phi}: R^n\longrightarrow R^n$ be defined by
$\overline{\phi}(c_0, c_1, \cdots, c_{n-1})=(c_0, (1+2u)c_1, (1+2u)^2c_2, \cdots, (1+2u)^{n-1}c_{n-1}).$  Then it is easy to prove the following lemma.

\hspace*{-0.6cm}{\bf Lemma 3.9} $C$ is a cyclic code over $R$ of odd length $n$ if and only if $\overline{\phi}(C)$ is a $(1+2u)$-constacyclic code of length $n$ over $R$.\\

Similarly to Theorem 4.8 in \cite{22}, we characterize $(1+2u)$-constacyclic codes over $R$ of odd length using the isomorphism $\phi$ as follows.\\
{\bf Proposition 3.10}
Let $n$ be odd and $C$ be a $(1+2u)$-constacyclic code of length $n$ over $R$. Then
$C$ is an ideal in $R[x]/\langle x^n-(1+2u)\rangle$ generated by
\begin{eqnarray*}
  C &=& \langle u_1^{\prime}(\tilde{x})(v_1'(\tilde{x})+2)+(1+u)u_2'(\tilde{x})(v_2'(\tilde{x})+2), (1+u)u_3'(\tilde{x})(v_3'(\tilde{x})+2)\rangle,
\end{eqnarray*}
where $\tilde{x}=(1+2u)x$, and $u_i'(x),v_i'(x),w_i'(x)$ are monic, pairwise coprime polynomials in $\mathbb{Z}_4[x]$ such that $x^n-1=u_i'(x)v_i'(x)w_i'(x),$ $i\in\{1,2,3\}$.\\

Before stating our next result about a class of constacyclic codes with a special generator polynomial, we need the following lemma.\\
{\bf Lemma 3.11} Let $n$ be odd. Suppose $C$ is a code over $R$ generated by $\langle u_1(x)(v_1(x)+2), uu_2(x)(v_2(x)+2)\rangle$.  Then $C$ is a cyclic code of length $n$ over $R$, where $ u_i(x),v_i(x),w_i(x)$ are monic, pairwise coprime polynomials in $\mathbb{Z}_4[x]$ such that $x^n-1=u_i(x)v_i(x)w_i(x)$, $i\in\{1,2\}$.

\begin{proof} By the assumption and $u\cdot[uu_2(x)(v_2(x)+2)]=u_2(x)(v_2(x)+2)\in C$, then $C$ can be expressed as $$C=\langle u_1(x)(v_1(x)+2), u_2(x)(v_2(x)+2)\rangle=\langle u_1(x)v_1(x), 2u_1(x), u_2(x)v_2(x), 2u_2(x)\rangle.$$

Suppose $d_1(x)=$gcd$(u_1(x)v_1(x),u_2(x)v_2(x))$ and $d_2(x)=$gcd$(u_1(x),u_2(x))$. Obviously, we have $\langle d_1(x),2d_2(x)\rangle \subseteq C$. On the other hand, $C=u_1(x)v_1(x)R[x]+u_2(x)v_2(x)R[x]+2u_1(x)R[x]+2u_2(x)R[x]$. However, $2u_1(x)R[x]+2u_2(x)R[x]=2d_2(x)R[x]$ and
$u_1(x)v_1(x)R[x]\\+u_2(x)v_2(x)R[x]=d_1(x)R[x].$ Hence, $C\subseteq \langle d_1(x),2d_2(x)\rangle$, i.e. $C= \langle d_1(x),2d_2(x)\rangle$, where $d_2(x)|d_1(x)|x^n-1$. According to the proof of Theorem 7.26 in \cite{W}, we have $C=\langle d_1(x)+2d_2(x)\rangle$. Thus $C$ is a cyclic code.
\end{proof}

Using the isomorphism $\phi$ and the above lemma, we characterize $(1+2u)$-constacyclic codes with special generator polynomial over $R$ of odd length as follows.\\
{\bf Theorem 3.12} Let $n$ be odd and $C$ be a $(1+2u)$-constacyclic code of length $n$ over $R$. Then
$C$ is an ideal in $R[x]/\langle x^n-(1+2u)\rangle$ generated by $C=\langle u_1(\tilde{x})(v_1(\tilde{x})+2), uu_2(\tilde{x})(v_2(\tilde{x})+2)\rangle$, where $\tilde{x}=(1+2u)x$, and $u_i(x),v_i(x),w_i(x)$ are monic, pairwise coprime polynomials in $\mathbb{Z}_4[x]$ such that $x^n-1=u_i(x)v_i(x)w_i(x)$, $i\in\{1,2\}$.\\

We can write the generators of a $(1+2u)$-constacyclic code given by the above theorem in the form $C=\langle g_1(\tilde{x}),ug_2(\tilde{x})\rangle$, where $\tilde{x}=(1+2u)x$.\\
{\bf Remark 3.13} In fact, according to Proposition 3.10 and the proof of Lemma 3.11, $C=\langle u_1'(\tilde{x})(v_1'(\tilde{x})+2)+(1+u)u_2'(\tilde{x})(v_2'(\tilde{x})+2), (1+u)u_3'(\tilde{x})(v_3'(\tilde{x})+2)\rangle,$ and if we set $u_2'(\tilde{x})(v_2'(\tilde{x})+2)=0, u_3'(\tilde{x})=u_1'(\tilde{x})=d_2(\tilde{x}), v_3'(\tilde{x})=v_1'(\tilde{x})=\frac{d_1(\tilde{x})}{d_2(\tilde{x})}$, then $C=\langle u_1'(\tilde{x})(v_1'(\tilde{x})+2)\rangle=\langle d_1(\tilde{x})+2d_2(\tilde{x})\rangle=\langle u_1(\tilde{x})(v_1(\tilde{x})+2), uu_2(\tilde{x})(v_2(\tilde{x})+2))\rangle,$ which means Proposition 3.10 includes Theorem 3.12 as a special case.\\

There is a special permutation of $\mathbb{Z}_4^{2n}$, called Nechaev permutation, which turns out to be useful in studying cyclic codes over $\mathbb{Z}_4$. It is defined as follows.\\
{\bf Definition 3.14}
Let $n$ be odd and let $\tau$ be the following permutation $\tau=(1, n+1)(3, n+3)\cdots(2i+1, n+2i+1)\cdots(n-2, 2n-2)$ on $\{0, 1, \cdots, 2n-1\}$. The Nechaev permutation is the permutation $\pi$ defined by $\pi(c_0, c_1, \cdots, c_{2n-1})=(c_{\tau(0)}, c_{\tau(1)}, \cdots, c_{\tau(2n-1)}).$\\

\hspace{-0.6cm}{\bf Proposition 3.15}
Let $\overline\phi$ be defined as above. If $\pi$ is the Nechaev permutation and $n$ is odd, then $\Phi \overline\phi=\pi\Phi.$

\begin{proof} Let $r=(r_0, r_1, \cdots, r_{n-1})\in R^n$ where $r_i=a_i+b_iu, 0\leq i\leq n-1.$ Since
$\overline\phi(r)=(r_0, (1+2u)r_1, (1+2u)^2r_2, \cdots, (1+2u)^{n-1}r_{n-1})$, $(\Phi \overline\phi)(r)=(b_0, 2a_1+b_1, b_2, 2a_3+b_3, \cdots, b_{n-1}, \\2a_0+b_0, b_1, 2a_2+b_2, b_3, \cdots, b_{n-2}, 2a_{n-1}+b_{n-1}).$
On the other hand, since $\Phi(r)=(b_0, b_1, \cdots, \\b_{n-1}, 2a_0+b_0, 2a_1+b_1, \cdots, 2a_{n-1}+b_{n-1})$, $(\pi\Phi)(r)=\pi(b_0, b_1, \cdots, b_{n-1}, 2a_0+b_0, 2a_1+b_1, \cdots, 2a_{n-1}+b_{n-1})=(b_0, 2a_1+b_1, b_2, 2a_3+b_3, \cdots, b_{n-1}, 2a_0+b_0, b_1, 2a_2+b_2, b_3, \cdots, b_{n-2}, \\2a_{n-1}+b_{n-1}).$
Thus $\Phi \overline\phi=\pi\Phi$.
\end{proof}
\hspace{-0.6cm}{\bf Corollary 3.16}
Let $\pi$ be the Nechaev permutation and $n$ be an odd number. If $\chi$ is the Gray image of a cyclic code over $R$, then $\pi(\chi)$ is a cyclic code.
\begin{proof} Let $\chi$ be such that $\chi=\Phi(C)$, where $C$ is a cyclic code over $R$. According to Proposition 3.15, we have $(\Phi \overline\phi)(C)=(\pi\Phi)(C)=\pi(\chi).$ By Corollary 3.9, we know that $\overline\phi(C)$ is a $(1+2u)$-constacyclic code. Therefore, by Corollary 3.2, $(\Phi \overline\phi)(C)=\pi(\chi)$ is a cyclic code.
\end{proof}

\section{Computational results}
In this section, based on Theorem 3.12,  we present the numerical results of a computer search on $(1+2u)$-constacyclic codes over $R$ and their $\mathbb{Z}_4$-images for some odd lengths. The computations are carried out using Magma software \cite{B}. We list both the minimum Lee weights and the minimum Euclidean weights for
$\mathbb{Z}_4$-images of the codes.  Recall that the Lee weights of $0, 1, 2, 3$ are, respectively, $0, 1, 2,1$ and the Euclidean weights of $0, 1, 2, 3$ are, respectively, $0,1,4,1$. We define the Lee and Euclidean weights of an element of $z=a+ub\in R$ as $w_L(z)=w_L(\Phi(z))=w_L(b, 2a+b)$ and $w_E(z)=w_E(\Phi(z))=w_E(b, 2a+b)$.\\
\hspace*{0.5cm}In Table 1 and Table 2, we obtain some  cyclic codes \textbf{with improved parameters}  over ${\mathbb Z}_4$ with respect to Lee weight (Euclidean weight) which are obtained from the $(1+2u)$-constacyclic codes $\left<{ g}_1(\tilde{x}),u{ g}_2(\tilde{x})\right> $ over $R$. That is, these codes have better parameters (larger minimum distance than the comparable codes) than the ones in \cite{D}. In some cases, a code of given size does not exist in \cite{D}. Those codes are also considered new. The first column is the length $n$ of the code over $R$, the second and third columns are the coefficients of generator polynomials written from high to low order (for example, the polynomial $x^4+3x^3+2x^2+1$ is represented by $13201$), where  ${g}_i(\tilde{x})=g_i((1+2u)x), i=1,2$ and the fourth column gives the parameters of the Gray images with respect to minimum Lee distance $d_L$ (minimum Euclidean distance $d_E$).

The codes with asterisk ($^*$) have the property that their binary images are linear and they are best known binary linear codes in  \cite{Data2}. The $(1+2u)$-constacyclic codes over $R$ have better $\mathbb{Z}_4$-parameters than cyclic codes \cite{22}. For example, in Table 1, for $n=7$,  the codes of length $14$ with minimum Lee distances $8,6,4$ have sizes $2^{10},2^{12},2^{19}$ respectively, while the codes in Table 3 of \cite{22}  only have sizes $2^6, 2^{10},2^{18}$ respectively. Also, in Table 2, for $n=7$,  the codes of length $14$ with minimum Euclidean distances $16,8,4$ have sizes $2^{9},2^{13},2^{19}$ respectively, while the codes in Table 3 of \cite{22}  only have sizes $2^6, 2^{12},2^{18}$ respectively.

Table 3 is a correction to the table of cyclic codes of length 7 over $R$ and their $\mathbb{Z}_4$-images that have length 14 published in \cite{22}. The entries in the table in the published article are erroneous. We present the corrected table here.
\begin{table}\label{table:1}
\caption{Some ${\mathbb Z}_4$ cyclic codes with \textbf{improved parameters} from $(1+2u)$-constacyclic codes over $R$ with respect to Lee weight}
$$\begin{array}{|c|c|c|c|c|}
\hline
n&  g_1(x)&g_2(x)& \text{Parameters of } {\mathbb Z}_4 \ {\rm image}\\
\hline
3&0&22&(6,4^02^2,8_L)^{\ast}\\
3&111&0&(6,4^12^1,6_L)^{\ast}\\
3&0&13&(6,4^22^3,4_L)^{\ast}\\
5&11111& 0&(10,4^12^1,10_L)\\

5&0&    11&(10,4^02^4,8_L)\\
5&0&    13&(10,4^42^5,4_L)^{\ast}\\
7&0&22202&(14,4^02^3,16_L)^{\ast}\\

7&1113133&2022&(14,4^12^1,14_L)\\
7&1113133& 1011&(14,4^12^4,12_L)^{\ast}\\
7& 3121&1113313&(14, 4^4 2^2, 8_L)\\
7&3121 &20222&(14,4^42^4,6_L)\\
7&0 &33&(14,4^62^7,4_L)^{\ast}\\
9&0&22022022&(18, 4^0 2^2, 24_L)^{\ast}\\
9&33033033&2002002&(18, 4^2 2^3, 12_L)\\
9&33033033&222&(18, 4^2 2^8, 8_L)\\
9&0&31&(18, 4^8 2^9, 4_L)\\

11& 33333333333&0&(22, 4^1 2^1, 22_L)\\

11&0& 22&(22, 4^02^{10}, 8_L)\\
11&0&13&(22, 4^{10}2^{11}, 4_L)\\

13&0&3333333333333&(26, 4^1 2^1, 26_L)\\
13&0&22&(26, 4^1 2^{12}, 8_L)\\
13&0&13&(26, 4^{12} 2^{13}, 4_L)\\
15&0&200020222&(30, 4^0 2^7, 20_L)\\
15&0&202022&(30, 4^0 2^{10}, 16_L)\\
15&0&10011&(30, 4^0 2^{11}, 12_L)\\
15&113212223&0&(30,4^7 2^{11}, 10_L)\\
15&113212223&1131023&(30, 4^{10 }2^{14}, 8_L)\\
15&13201&13201&(30, 4^{11} 2^{11}, 6_L)\\
15&1131023&1131023&(30, 4^9 2^{13}, 6_L)\\
15&0&11&(30, 4^{14}2^{15}, 4_L)\\
\hline
\end{array}$$
\end{table}

\begin{table}\label{table:2}
\caption{Some ${\mathbb Z}_4$ cyclic codes with \textbf{improved parameters} from $(1+2u)$-constacyclic codes over $R$ with respect to Euclidean weight}
$$\begin{array}{|c|c|c|c|}
\hline
n&  g_1(x)&g_2(x)& \text{Parameters of } {\mathbb Z}_4 \ {\rm image}\\
\hline
3&0&22&(6,4^02^2,16_E)\\
3&0&2002&(6,4^02^3,8_E)\\
3&0&113&(6,4^12^3,6_E)\\

5&0&  22&(10,4^02^4,16_E)\\
5&11111&11111&(10,4^12^1,10_E)\\
5&0& 13&(10,4^42^5,8_E)\\
7&0& 11101&(10,4^02^3,32_E)\\
7& 0&22202&(14,4^02^4,24_E)\\
7&31101& 1111111&(14, 4^3 2^3, 16_E)\\

7&1113313& 1113133&(14, 4^1 2^4, 14_E)\\
7&& 32111&(14, 4^3 2^7, 8_E)\\
7&0 & 3121&(14, 4^4 2^7, 6_E)\\
7&0 & 33&(14, 4^6 2^7, 4_E)\\
9&0&22022022&(18, 4^0 2^2, 48_E)\\
9&0&2002002&(18, 4^0 2^3, 24_E)\\
9&0&22&(18, 4^0 2^8, 16_E)\\
9&33033033&222&(18, 4^2 2^8, 12_E)\\

9&0&1001003&(18, 4^3 2^9, 6_E)\\
9&0&31&(18, 4^8 2^9, 4_E)\\

11& 33333333333&0&(22, 4^1 2^1, 22_E)\\

11&0& 22&(22, 4^02^{10}, 16_E)\\
11&0&13&(22, 4^{10}2^{11}, 4_E)\\

13&0&3333333333333&(26, 4^1 2^1, 26_E)\\
13&0&22&(26, 4^1 2^{12}, 16_E)\\
13&0&13&(26, 4^{12} 2^{13}, 4_E)\\
15&0&200020222&(30, 4^0 2^7, 40_E)\\
15&0&202022&(30, 4^0 2^{10}, 32_E)\\
15&0&10011&(30, 4^0 2^{11}, 24_E)\\
15&0&22&(30, 4^02^{14}, 16_E)\\
15&113212223&0&(30, 4^7 2^{11}, 10_E)\\
15&113212223&1131023 &(30, 4^{10} 2^{14}, 8_E)\\
15&0&30211&(30, 4^{11}2^{15}, 6_E)\\
15&0&11&(30, 4^{14}2^{15}, 4_E)\\
\hline
\end{array}
$$
\end{table}

\begin{table}
\noindent\caption{\small{Correction to \cite{22}: Some cyclic codes of length 7 with $\mathbb{Z}_4$-images}}
\begin{center}
\begin{tabular}
{|c|c|c|c|} \hline ${g_1(x)}$ & \ \ \  ${g_2(x)}$\ \ & ${g_3(x)}$  &  \text{Parameters of } $\mathbf{\Z_4}$ \text{image}
\\\hline
$20222$ & $20222$ & $20222$ &$(14,4^02^6,8)$
\\ $2202$ & $2222222$ & $2222222$ &$(14,4^02^8,6)$
\\ $22$ & $20222$ & $20222$ &$(14,4^02^{12},4)$
\\ $2^7$ & $3^7$ & $3^7$ &$(14,4^12^1,7)$
\\ $2202$ & $3^7$ & $3^7$ &$(14,4^12^7,6)$
\\ $1113313$ & $3^7$ & $3^7$ &$(14,4^22^6,6)$
\\ $22$ & $32133$ & $32133$ &$(14,4^32^9,4)$
\\ $2$ & $3^7$ & $3121$& $(14,4^42^{10},2)$
\\ $11301$ & $20222$ & $20222$& $(14,4^62^6,4)$
\\ $11$ & $2^7$ & $2^7$& $(14,4^{12}2^2,2)$

\\ $20222$ & $20222$ & $20222$ &$(14,4^02^6,16_E)$

\\ $20222$ & $2^7$ & $2^7$ & $(14,4^02^7,12_E)$
\\ $22$ & $20222$ & $20222$& $(14,4^02^{12},8_E)$
\\ $20222$ & $3^7$ & $3^7$ &$(14,4^12^{6},12_E)$
\\ $2^7$ & $2^7$ & $32133$ &$(14,4^{3}2^{1},12_E)$
\\ $20222$ & $32133$ & $32133$ &$(14,4^{3}2^{3},8_E)$
\\ $20222$ & $3121$ & $3^7$ &$(14,4^{4}2^{0},12_E)$
\\ $22$ & $3^7$ & $32133$ &$(14,4^{4}2^{9},4_E)$
\\ $3^7$ & $3^7$ & $32133$ &$(14,4^{5}2^{0},7_E)$
\\ $3121$ & $3^7$ & $33123$ &$(14,4^{11}2^{0},3_E)$
\\\hline
\end{tabular}
\end{center}
\end{table}

\section{Conclusion}
\hspace*{0.6cm}This article is devoted to investigating some properties of $(1+2u)$-constacyclic codes over $R=\Z_4+u\Z_4,$ where $u^2=1$. We present many examples of $(1+2u)$-constacyclic codes over $R$ whose $\mathbb{Z}_4$ images are $\mathbb{Z}_4$-cyclic codes with improved parameters according to the online database \cite{D}. It is worth exploring properties of constacyclic codes over the other rings of order 16 \cite{MS}, and examine whether they produce new linear codes over $\mathbb{Z}_4.$

\section{Acknowledgement}
\hspace*{0.6cm} This research is
supported by National Natural Science Foundation of China (61672036), the Open Research Fund of National Mobile Communications Research Laboratory, Southeast University (2015D11), Technology Foundation for Selected Overseas
Chinese Scholar, Ministry of Personnel of China (05015133) and Key projects of support program for outstanding young talents in Colleges and Universities (gxyqZD2016008).

\end{document}